\documentstyle[12pt]{article}
\begin{document}
\begin{titlepage}
\begin{flushright}
{\normalsize\bf HU-SEFT R 1993-10}
\end{flushright}
\vspace{1cm}
\begin{center}
\vspace*{1.0cm}
{\Large q-Deformed Path Integral}
\vskip 2.0cm
by
\vskip 0.5 cm

{\large M. Chaichian}\renewcommand{\thefootnote}{*}
\footnote{High Energy Physics
Laboratory, Department of Physics
and
Research Institute for
High Energy Physics, P.O.Box 9 (Siltavuorenpenger
20 C), SF-00014, University of Helsinki, Finland;
E-mail address: chaichian@finuhcb.helsinki.fi} $and$
{\large A.P.Demichev}\renewcommand{\thefootnote}{\dagger}\footnote{Nuclear
Physics Institute, Moscow State University, 119899, Moscow, Russia; E-mail
address: demichev@compnet.msu.su}
\end{center}
\vspace{3 cm}
\begin{abstract}
\normalsize

Using differential and integral calculi on the quantum plane which are
invariant with respect to quantum inhomogeneous Euclidean group
$E(2)_q$, we construct path integral representation for the quantum
mechanical evolution
operator kernel of q-oscillator.

\end{abstract}
\end{titlepage}

\section{Introduction}
Deformations of different groups and algebras \cite{Drinf}-\cite{Wor}
has attracted great attention during the last few years.
These mathematical objects
originate in quantum inverse scattering method \cite{Fadd} and have
found many interesting and important physical applications. A partial list
of possible applications in quantum field theory and particle physics
includes:
{\it i)} q-deformations of the space-time symmetry groups
and Lie algebras (see e.g. \cite{SWZ} - \cite{CD} and refs. therein)
with the hopes to obtain naturally regularized field theory;
{\it ii)}  attempts in q-deformations of internal (gauge) symmetries
\cite{AV} - \cite{Madj};
{\it iii)}  q-deformations of Heisenberg algebra of raising and
lowering operators (see e.g. \cite{MF} - \cite{Kempf} and refs.
therein).

In this paper we deal with the last topic and consider the problem of
the path integral representation of the quantum mechanical
evolution operator kernel for
q-deformed oscillator, the corresponding raising and lowering
operators ${\bf b^+}$ and $\bf b$ obeying the
following commutation relation (CR)
\begin{equation}
{\bf bb^+} - q^2{\bf b^+b}=1, \qquad q \in {\bf R}\ .          \label{1}
\end{equation}
There
are two possible way to consider the relation (\ref{1}). In the first
case one considers ${\bf b,b^+}$-operators as constructed from usual
canonical variables. This leads to Macfarlane's representation \cite{MF}
of (\ref{1}),
\begin{eqnarray}
{\bf b}^+=\bar\alpha\left[ e^{2isx}-e^{is\partial}e^{isx} \right]\
, \nonumber \\
{\bf b}=\alpha\left[ e^{-2isx}-e^{-isx}e^{is\partial}\right] \ ,
\label{2}
\end{eqnarray}
where $q=e^{-s^2},\ \alpha\bar\alpha=(1-q^2)^{-1} $. As is seen
from (\ref{2}) one must identify
coordinates $x$ and $x+2\pi s$ so that configuration space of
q-oscillator is compact and topologically equivalent to a circle.
The basic CR is the canonical one and CR (\ref{1}) has an auxiliary meaning.
Correspondingly, one can represent operators ${\bf b},{\bf b}^+$ in terms of
the
ordinary differential operators and operators of multiplication by
coordinate.

In the second case one considers CR (\ref{1}) as the basic relation
and represents \cite{PWor}, \cite{Kempf}
 raising and lowering q-operators in terms of differential
operators and coordinates \cite{WZ} on quantum plane .

In accordance with those views on the nature of CR (\ref{1}) there exist
two approaches in the construction of the path integral. The first approach
was considered  carefully and in detail in \cite{Shab}. As a result
the usual (bosonic) path integral on topologically non-trivial
(non-deformed) phase space was obtained, characteristic size of
the latter being
connected with q-parameter. Two notes are in order: {\it i)} even the
simplest Hamiltonian in term of $\bf b$-operators, namely
\begin{equation}
{\bf H}=\omega {\bf b^+b}\ ,                   \label{3}
\end{equation}
has a very complicated non-polynomial form in term of canonical
variables $x$ and $p$. So the path integral can not be calculated
explicitly (has non-Gaussian form); {\it ii)} it is
incomprehensible to see how this
formulation can be used for construction of q-deformed quantum field
theory.

It is well known that for quantization of the field theories path
integral in holomorphic representation is most suitable (see e.g.
\cite{FS}). From the latter point of view and taking into account
the analogy with fermionic (Berezin) path integral, it would be
desirable to construct path integral over "classical" analog of
operators ${\bf b},{\bf b}^+$, that is variables $z, \bar z$, satisfying the CR
\begin{equation}
z\bar z - q^2\bar zz=0\ ,          \label{4}
\end{equation}
in other words over variables on a quantum plane \cite{Man}, \cite{WZ}. Such an
attempt was made in \cite{BF}. However, the authors of \cite{BF}
aimed to construct the evolution operator for modified Schr{\"o}dinger
equation
\begin{equation}
iD_q\Psi (t) = {\bf H}({\bf b,b^+})\Psi (t) \ ,           \label{5}
\end{equation}
where
\begin{equation}
D_q f(t)={{f(q^2t)-f(t)}\over{t(q^2-1)}}       \label{6}
\end{equation}
is the q-derivative. First of all we note that such a choice of time
evolution equation is completely independent of q-CR relation
(\ref{1}),  that  is  one  could  assume  q-deformed  form  of
Scr\"{o}dinger
equation for usual oscillator and vice versa. Operator solution of
eq. (\ref{5}) has the following form \cite{BF}:
\begin{eqnarray}
\Psi (t)={\bf U}(t,t_0)\Psi(t_0)\ ,\\                 \label{7}
{\bf U}(t,t_0)=\exp_{q^{-1}}\{-it_0{\bf H}\}\exp_q\{it{\bf H}\}\ , \label{8}
\end{eqnarray}
where
\begin{equation}
\exp_qA=\sum_{n=0}^\infty \frac{A^n}{[n]!}       \label{9}
\end{equation}
is q-deformed exponential \cite{Ex}, and
\begin{equation}
[X]={{q^{2X}-1}\over{q^2-1}}\ .
\end{equation}
Due to the property of the q-exponents
$$
\exp_q\{A\}\exp_{q^{-1}}\{-A\}=1\ ,
$$
The operator $\bf U$ has the necessary properties of an evolution operator:
\begin{eqnarray}
{\bf U}(t,t)=1\ ,  \nonumber  \\
{\bf U}(t_1,t_2){\bf U}(t_2,t_3)={\bf U}(t_1,t_3)\ ,         \label{11}
\end{eqnarray}
but, unfortunately, $\bf U$ is not unitary operator
$$
{\bf U^+}(t,t_0){\bf U}(t,t_0)=exp_q\{-it{\bf H}\}
exp_{q^{-1}}\{it_0{\bf H}\}
exp_{q^{-1}}\{-it_0{\bf H}\}exp_q\{it{\bf H}\}\neq 1\ ,\qquad
\mbox{if} \ \ q\neq 1\ .
$$
This contradicts the basic principle of quantum mechanics
(probability interpretation etc.).
Note that from the point of view of Heisenberg equation of motion the
absence of unitarity for q-evolution was shown in \cite{CE}. So we
will not proceed in this way, but consider the usual
(non-deformed in time) Schr{\"o}dinger equation.
We must add also that in a part of their paper the authors of \cite{BF}
consider $q^2$ as pure phase
\renewcommand{\thefootnote}{1}
\footnote{This pure phase in the case of $q^2$
 a root of unity of order $k$ has been associated in [20] with the
very interesting possibility of obtaining an extension of Berezin calculus as
$z^k=0,\bar{z}^k=0$. We would like to notice that with the case of deformation
parameter being a pure phase, one ought to use other
 $q$-oscillator operators: $a=q^{-N/2}b,a^+=b^+q^{N/2}$. Then
the whole construction of Bargmann-Fock representation has to be revised
in this case.}: $q^2=e^{is}, s \in \bf R$. In this case
operators $\bf b^+$ and $\bf b$ can not be conjugated to each other
since one would obtain after conjugation of (\ref{1}) meaningless
relation: ${\bf b^+b}=0$. Throughout a paper we shall consider $q \in \bf R$.
Most of the useful formulas from \cite{BF}, however, remain valid
also for this case. The central point of construction in \cite{BF} is the
deformed Bargmann-Fock representation (BFR) of (\ref{1}) in the space
of antianalytic functions on the q-plane. The differential calculus
used for construction of BFR is not invariant with respect to any
deformed $GL(N)_q$ group (in contrast to the author's statement) and
part of CR for derivatives and coordinates was chosen arbitrarily. We
will show that the latters are in fact fixed from the requirement of invariance
with respect to the quantum Euclidean group $E(2)_q$. The final formula
of \cite{BF} is the discrete approximation for path integral or, in
other words, convolution of finite number of (nonunitary) evolution
operators.

The aim of the present paper is to clarify the differential
calculus used for the construction of BFR and to obtain the path integral
in holomorphic representation for q-oscillator in full analogy with
usual (bosonic and fermionic) case.

We have to mention also the papers \cite{Jur}, where deformed path
integral was considered for q-group coherent states using Jackson
integral \cite{Ex} over usual (commuting) variables.
Such form of path integral can not be used for possible further
applications in quantum field theory.

The content of the paper is the following. In next section we
introduce $E(2)_q$-invariant differential calculus on a quantum plane
which is necessary for construction of BFR. Sect.3 is devoted to
derivation of the q-deformed path integral. Sect.4 contains the
Conclusion.

\vskip 1.0 cm

\section{$E(2)_q$-invariant differential calculus
and Bargmann-Fock representation}

Usually in the problems concerning q-oscillators the well
developed theory of q-deformed semisimple matrix group is used. They are
useful in two aspects. Firstly, one can consider the couple of bosonic
$\{{\bf b}_i,{\bf b}_i^+\}_{i=1}^N$ and fermionic
$\{{\bf f}_j,{\bf f}_j^+\}_{j=1}^N$
oscillators with CR which are invariant with respect to quantum group
or supergroup \cite{CKL},\cite{Kempf}. Secondly, if one tries to construct
the BFR of (\ref{1}) or its multioscillator generalization, i.e. to
maintain the correspondence
${\bf b}^+_i\ \rightarrow\ \bar z_i,\ \ {\bf b}_i\
\rightarrow \bar\partial_i$
(here $\bar z_i,\ \bar\partial_i$ are coordinates and derivatives on
q-hyperplane), one needs the CR between coordinates and derivatives,
in other words, the q-differential calculus. Usually, one develops the
$GL(N)_q$-invariant differential calculus \cite{WZ} (see also
\cite{Schirr} for multiparametric deformation case).

Let us
consider now the basic case of just one (bosonic) oscillator. In this
case we have to use the 2-dimensional q-plane with coordinates $z, \bar
z$, satisfying CR (\ref{4}), and the corresponding derivatives $\partial,
\ \bar\partial$. As is well known, BFR is constructed in the space of
antianalytic functions $f(\bar z )$ (the coordinates $z$ and $\bar z$
are conjugated to each other). Hence, the group of invariance can not mix
$z$ and $\bar z$ (can not be $GL(2)_q$) and the homogeneous
transformations  has the following matrix form:
\begin{equation}
\left( \begin{array}{c}
z^\prime \\
\bar z^\prime
\end{array}  \right) =
\left( \begin{array}{cc}
u & 0 \\
0 & \bar u
\end{array}  \right)
\left( \begin{array}{c}
z \\
\bar z
\end{array}  \right)\ .                              \label{12}
\end{equation}
It is easy to see that such transformations do not fix differential
calculus on the q-plane. From the other hand, derivatives are
connected with translations and, hence, differential calculus are
connected with inhomogeneous groups. So it is natural to demand
invariance of the calculus with respect to inhomogeneous
generalization of (\ref{12}),  i.e.  with  respect  to  deformed
Euclidean
group $E(2)_q$. The theory of deformation of non-semisimple
inhomogeneous groups is not so well developed as the theory of
semisimple ones (see, however, \cite{SWW}, \cite{CD2}).
Fortunately, the 2-dimensional case is the simplest one and has been
considered in a number of papers \cite{VK}.

$E(2)_q$-transformations have the form
\begin{eqnarray}
z\ \longrightarrow z^\prime = uz+t\ , \nonumber \\
\bar z\ \longrightarrow \bar z^\prime = \bar u\bar z+\bar t\ .
                                                \label{13}
\end{eqnarray}
The CR for the group parameters are
\begin{eqnarray}
\bar u u = u\bar u = 1\ , \nonumber \\
t\bar t =q^2 \bar t t\ , \nonumber \\
u t =q^2 t u\ , \nonumber \\
\bar u t =q^{-2}t \bar u\ .                    \label{14}
\end{eqnarray}
Other CR can be obtained after involution of (\ref{14}).
Comultiplication $\Delta$ for group elements is defined as follows
\begin{equation}
\begin{array}{cc}
\Delta (u) = u \otimes u\ , & \Delta (\bar u) = \bar u \otimes \bar u\ , \\
\Delta (t) = u \otimes t + t \otimes {\bf 1}\ , &
\Delta (\bar t) = \bar u \otimes \bar t + \bar t \otimes {\bf 1}\ .
\end{array}                                            \label{15}
\end{equation}
This comultiplication can be expressed in the usual matrix form (cf.
\cite{Drinf} - \cite{Wor}),
$$
\Delta  ({\bf M}_{ij}) ={\bf M}_{ik}\otimes {\bf M}_{kj}\ ,
$$
if one writes the transformations (\ref{13}) in the matrix projective
form
\begin{equation}
\left( \begin{array}{c}
z^\prime \\
\bar z^\prime \\
\bf 1
\end{array} \right) =
{\bf M}
\left( \begin{array}{c}
z \\
\bar z \\
\bf 1
\end{array} \right) \equiv
\left( \begin{array}{ccc}
u & 0 & t \\
0 & \bar u & \bar t \\
0 & 0 & {\bf 1}
\end{array} \right)
\left( \begin{array}{c}
z \\
\bar z \\
\bf 1
\end{array} \right)                   \ .           \label{16}
\end{equation}
This form is also convenient for writing the antipode:
$$
\left( \begin{array}{ccc}
u & 0 & t \\
0 & \bar u & \bar t \\
0 & 0 & {\bf 1}
\end{array} \right)^{-1} =
\left( \begin{array}{ccc}
\bar u & 0 & -\bar u t \\
0 & u & -u\bar t \\
0 & 0 & {\bf 1}
\end{array} \right)\ ,
$$
and counity: $\epsilon ({\bf M}_{ij})=\delta_{ij}$.
To develop the differential calculus, we introduce the q-differentials
$dz,\ d\bar z$ which are transformed homogeneously: $dz\ \rightarrow
udz,\ d\bar z\ \rightarrow \bar u d \bar z$ and require that CR for
the differentials and coordinates to be invariant with respect to
$E(2)_q$-transformations. Let us consider as an example the CR for
$z$ and $d\bar z$. The general form of CR is the following:
\begin{equation}
zd\bar z = A_1(d\bar z) z + A_2(dz)z + A_3(d\bar z)\bar z +
A_4(dz)\bar z\ ,                                      \label{18}
\end{equation}
where $A_i$ are constants to be defined from invariance condition.
After $E(2)_q$-transformation of (\ref{18}), one has
\begin{eqnarray*}
zd\bar z + t\bar udz = A_1(d\bar z)z + A_2u^2(dz)z +A_3\bar u^2(d\bar
z)\bar z+ A_4(dz)\bar z  \\
+ A_1\bar u td\bar z + A_2utdz +A_3\bar u\bar td\bar z
+ A_4u\bar tdz\ .
\end{eqnarray*}
Comparing both sides of the equation, one obtains
$$
A_1 = q^2\ ,        $$
$$
A_2=A_3=A_4=0\ .    $$
Hence
\begin{equation}
zd\bar z = q^2d\bar z z\ .                         \label{19}
\end{equation}
Introducing external differential $d$ with the usual property $d^2=0$, one
immediately derives from (\ref{19}) the CR for differentials:
$$
dzd\bar z = - q^2 d\bar zdz   \ .     $$
The CR for other pairs of coordinates and differentials are obtained
analogously.  Now one can introduce q-derivatives $\partial , \
\bar\partial$ through the relation
$$
d=d\bar z\bar\partial + dz\partial\ ,              $$
and using the Leibnitz rule for external differential $d$ (but not for
derivatives) \cite{WZ}, one can derive the complete set of CR for
coordinates, differentials and derivatives:
\begin{equation}
\begin{array}{ccc}
z\bar z = q^2\bar z z \ , & \qquad & \partial\bar\partial = q^2
\bar\partial\partial\ , \\
\qquad & \qquad & \qquad \\
dzd\bar z= - q^2d\bar z dz\ , &  \qquad &(dz)^2=(d\bar z)^2=0\ , \\
\qquad & \qquad & \qquad \\
zd\bar z = q^2(d\bar z) z\ , & \qquad & \bar z dz = q^{-2}(dz)\bar
z\ ,\\
\qquad & \qquad & \qquad \\
zdz = q^{-2}(dz) z\ , & \qquad & \bar z d\bar z = q^2(d\bar z)\bar
z\ ,\\
\qquad & \qquad & \qquad \\
\partial z=1+q^{-2}z\partial\ , & \qquad &
\bar\partial\bar z=1+q^2\bar z\bar\partial\ , \\
\qquad & \qquad & \qquad \\
\partial\bar z=q^{-2}\bar z\partial\ , &  \qquad &
\bar\partial z=q^2 z\bar\partial\ , \\
\qquad & \qquad & \qquad \\
dz\bar\partial=q^{-2}\bar\partial dz\ , &  \qquad &
d\bar z\partial=q^2\partial d\bar z\ ,  \\
\qquad & \qquad & \qquad \\
d\bar z\bar\partial=q^2\bar\partial d\bar z\ , &  \qquad &
dz\partial=q^{-2}\partial dz\ .
\end{array}                              \label{20}
\end{equation}
These CR coincide with those in \cite{BF}. Thus we have shown that the
differential calculus used in\cite{BF} is not arbitrary and not fixed
by $GL(2)_q$-group but is defined by inhomogeneous Euclidean quantum
group. Note that it considerably differs from $GL(2)_{q,r}$-invariant
calculus developed in \cite{WZ}, \cite{Schirr}.

As is shown in \cite{BF} this differential calculus defines in turn
the integral calculus if one adopts the following postulates:
\begin{itemize}
\begin{enumerate}
\item
Normalization
$$
I_{00}= \int d\bar zdz\ e_q^{-\bar z z} \ = 1; $$
\item
Analog of the Stokes formula
$$
\int d\bar z dz\ \bar\partial f(\bar z,z)=\int d\bar z dz\ \partial
f(\bar z,z)=0\ ;$$
\item
Change of variables
$$
\int d\bar z dz\ e_q^{-a^2\bar z z}f(\bar z, z)=a^{-2}\int d\bar z dz\
e_q^{-\bar z z}f(a^{-1}\bar z,a^{-1}z)\ .  $$
\end{enumerate}
\end{itemize}
These postulates are sufficient to prove the following result,
\begin{equation}
I_{mn}=\int d\bar z dz\ e_q^{-\bar z z}z^n\bar z^m = \delta_{mn}
[n]!\ ,                                           \label{21}
\end{equation}
which is sufficient to compute the integral of any function which has
a power series expansion.
\vskip 1.0 cm

\section{Path integral for q-oscillators}

The BFR for CR (\ref{1}) is constructed in the Hilbert space $\cal H$
of antianalytic functions $f(\bar z)$ on the q-plane with scalar
product of the form
\begin{equation}
<g,f>=\int d\bar z dz\ e_q^{-\bar z z} \overline{g(\bar z)}f(\bar z) \ ,
                                               \label{22}
\end{equation}
so that the monomials
\begin{equation}
\psi ^n (\bar z)=\frac{\bar z^n}{\sqrt{[n]!}}        \label{23}
\end{equation}
form the orthonormal complete set of vectors in $\cal H$. Lowering
and raising operators are represented as derivative and
coordinate
\begin{equation}
{\bf b}^+=\bar z\ ,\qquad {\bf b}=\bar\partial\ ,           \label{24}
\end{equation}
and using the formulas
\begin{eqnarray}
\bar\partial e_q^{a\bar z z}=aze_q^{a\bar z z}\ , \nonumber \\
\partial e_q^{a\bar z z}=aq^{-2}\bar ze_q^{aq^{-2}\bar z z}\ ,
                                                     \label{25}
\end{eqnarray}
one can check that $b^+$ and $b$ are hermitian conjugated to each
other with respect to the scalar product (\ref{22}).

As usual the action of any operator $A$ in $\cal H$ can be
represented with the help of its kernel
\begin{equation}
({\bf A}f)(\bar z_1) = \int d\bar z_2 dz_2\ e_q^{-\bar z_2 z_2} A(\bar
z_1,z_2)f(\bar z_2)\ ,                            \label{26}
\end{equation}
where
\begin{equation}
A(\bar z_1,z_2)=\sum_{m,n} A_{mn}
\frac{\bar z_1^m}{\sqrt{[m]!}}\frac{z_2^n}{\sqrt{[n]!}}\ .   \label{27}
\end{equation}
Here one more pair of q-commuting coordinates is introduced. So we
have to define the CR for coordinates on different copies of
q-planes. In \cite{BF}the simplest choice was made and postulated
that coordinates on different q-planes commute with each other. In
the \cite{Kempf} CR are nontrivial and defined by R-matrix. From our
point of view the choice of the CR depends on concrete meaning of
different planes. In our case these q-planes will correspond to
different time slices in the process of time evolution. In the
continuous limit they become infinitesimally close to each other and
it would be quite unnatural if coordinates on them would commute. More
formally this argument can be expressed as follows. We assume that the
classical  analog  of  oscillator  operators  ${\bf b^+,b}$,  i.e. the
variables
$\bar z,z$, obey some Hamiltonian-like equation of motion of the
general form
$$
\frac{dz}{dt}=F(\bar z(t),z(t))\ ,       $$
$$
\frac{d\bar z}{dt}=\bar F(\bar z(t),z(t))\ .       $$
RHS of these equations have definite CR with $\bar z(t),\ z(t)$.
Hence LHS must have the same CR. It means that $\bar z(t+\Delta t),\
z(t+\Delta t)$ have the same CR with $\bar z(t),\ z(t)$ as the latters
have CR with each other. As a result, we postulate that any copies of
coordinates $\bar z_i,\ z_i \ (i=1,2,...)$ on q-planes have the
following CR:
\begin{eqnarray}
z_i\bar z_j=q^2\bar z_jz_i\ ,
\qquad \bar z_i\bar z_j=\bar z_j\bar z_i\ ,\nonumber \\
z_iz_j=z_jz_i\ ,                                          \label{28}
\end{eqnarray}
i.e. they do not depend on the indices which distinguish the copies.
The same is true for CR with derivatives and differentials. Note that
this is more similar to the fermionic case than the case of commuting
copies of coordinates.

Now we can express $A_{mn}$ through scalar product
\begin{equation}
A_{mn}=q^{2m(n+1)}<\psi _m \|{\bf A}\|\psi _n>\ .      \label{29}
\end{equation}
The next formula which is necessary for constructing the path
integral, is the convolution of operator kernels. Let us consider the
action of two operators on some function $f(\bar z)$ from $\cal H$,
\begin{equation}
{\bf A_2}{\bf A_1}=\int d\bar z_1 dz_1 e_q^{-\bar z_1z_1}A_2(\bar
z_2,z_1) \int d\bar z_0 dz_0 e_q^{-\bar z_0 z_0}A_1(\bar
z_1,z_0)f(\bar z_0)\ .                               \label{30}
\end{equation}
 From (\ref{20}) and (\ref{28}) it follows that in general $A_2(\bar
z_2,z_1)$ does not commute with $\exp_q\{-\bar z_0z_0\}$ and one can
not express the kernels of operator ${\bf A_2A_1}$ through those of $\bf
A_2$ and $\bf A_1$. For example, if
\begin{equation}
A_2(\bar z_2,z_1)\bar z_0z_0=q^k\bar z_0z_0A_2(\bar z_2,z_1) \label{31}
\end{equation}
for some integer $k$, one has
\begin{equation}
{\bf A_2A_1}f(\bar z_2)=\int d\bar z_0dz_0 exp_q\{-q^k\bar z_0z_0\}
\left[ \int d\bar z_1dz_1 exp_q\{-\bar z_1z_1\}
A_2(\bar z_2,z_1)A_1(\bar z_1,z_0)\right]f(\bar z_0)\ .       \label{32}
\end{equation}
Note, that in \cite{Kempf} part of CR for different copies of
coordinates are fixed just from the requirement of existence of the usual
convolution formula. It is easy to see from (\ref{30}) that in the
case of one oscillator this leads to commuting different coordinates.
As we argued above this is not appropriated from the physical point
of view. To avoid this problem, we restrict our consideration to the
operators with the kernels of the form
$$
A(\bar z_i,z_i)=A(\bar z_iz_i)     \ .        $$
Such kernels commute with $\bar z_mz_n$, so that the integer $k$ in
(\ref{31}) is zero and from (\ref{32}) one has the usual convolution
formula. This is analogous to even operators restriction in the
fermionic case. If one has some operator in the normal form, e.g.
the monomial
$$
{\bf M}_k=({\bf b}^+)^k{\bf b}^k           $$
for some integer $k$, one obtains from (\ref{27}),(\ref{29}) the
expression for integral kernel
\begin{equation}
M_k(\bar z z)=\sum^\infty_{n=0}q^{2(n+k)(n+k+1)-2nk}\ \frac{\bar z^n
z^n}{[n]!} \bar z^kz^k   \ .               \label{33}
\end{equation}
Using the relation
\begin{equation}
\bar z^n z^n=q^{-n(n-1)}(\bar z z)^n\ ,        \label{33a}
\end{equation}
the expression (\ref{33}) can be written in the form
\begin{equation}
M_k(\bar z z)=q^{2k(k+1)}\left(
\sum^\infty_{n=0}q^{n(n-1)}
\frac{\left( q^{2(k+2)}\bar z z \right)^n}{[n]!}\right)\bar z^kz^k\ .
                                                \label{34}
\end{equation}
The sum in (\ref{34}) is nothing but the second basic exponential
function \cite{Ex}, $\exp_{1/q}\{x\}$, so that the final expression for
integral kernel of normal operator is
\begin{equation}
M_k(\bar z z)=q^{2k(k+1)}\exp_{1/q}\{q^{2(k+2)}\bar z z\}\bar z^kz^k\ .
                                                    \label{35}
\end{equation}
According to our discussion in the Introduction we consider the usual
Schr{\"o}dinger equation
\begin{equation}
i\frac{d\ }{dt} \Psi (\bar z,t)={\bf H}({\bf b^+,b})\Psi(\bar z,t)\ ,
\label{36}
\end{equation}
with the simplest nontrivial Hamiltonian\renewcommand{\thefootnote}{2}
\footnote{It is obvious
that for q-deformed oscillator trivial "free"
Hamiltonian is proportional to particle number operator i.e.
$H_0 \sim N$ (remind that $b^+b=[N]$); the
spectrum of $ H_0$ is the equidistant one.}
\begin{equation}
{\bf H(b^+,b)}=\omega {\bf b^+b}                \ . \label{37}
\end{equation}
 Integral kernel for infinitesimal operator
$$
{\bf U}\approx
{\bf 1}-i{\bf H}\Delta t={\bf 1}-i\omega{\bf b^+b}\Delta t\ , $$
takes the form
\begin{equation}
U(\bar z z)=e_{1/q}^{q^4\bar z z}\left( 1-q^4e_q^{-q^4(1-q^2)\bar z
z} i\omega\bar z z\Delta t\right) \approx e_{1/q}^{q^4\bar z z}
exp\{-iH_{eff}\Delta t\} \ ,                          \label{38}
\end{equation}
where
$$
H_{eff}=  q^4e_q^{-q^4(1-q^2)\bar zz}\omega\bar z z  \ .  $$
Here we used the summation theorem for q-exponentials for commuting
arguments \cite{Ex},
\begin{equation}
e_q^Ae_{1/q}^B=e_q^{A+B}\ ,\qquad AB=BA\ .   \label{39}
\end{equation}
Now we can write the convolution of $K$ infinitesimal evolution
operator kernels
$$
U(\bar z_Kz_{K-1})*U(\bar z_{K-1}z_{K-2})*...*U(\bar z_1z_0)\  =  $$
$$
\int d\bar z_{K-1}dz_{K-1}...d\bar z_1dz_1 e_q^{-\bar z_{K-1}z_{K-1}}
...e_q^{-\bar z_1z_1}                         $$
\begin{equation}
\times e_{1/q}^{-q^4\bar z_Kz_{K-1}}...
e_{1/q}^{-q^4\bar z_1z_0}e^{-iH_{eff}(\bar z_Kz_{K-1})\Delta t}
...e^{-iH_{eff}(\bar z_1z_0)\Delta t}\ .             \label{40}
\end{equation}
Using the product representation for the second q-exponent \cite{Ex},
$$
e_{1/q}^x=\prod ^\infty_{r=0} \{1+xq^{2r}(1-q^2)\}\ ,      $$
and introducing $\Delta z_{K-l}$ as
$$
z_{K-l-1}=z_{K-l}-\Delta z_{K-l} \ ,                $$
one can write
$$
e_{1/q}^{q^4\bar z_K z_{K-1}}...e_{1/q}^{q^4\bar z_1 z_0}\ =\
\exp\left\{ \sum^{K-1}_{l=1}\ln\ e_{1/q}^{q^4\bar z_{K-l}
z_{K-l-1}}\right\}                                $$
$$
=\exp\left\{\sum_{l=1}^{K-1}\sum_{r=0}^\infty \ln(1+
q^{2r+4}(1-q^2)\bar z_{K-l}z_{K-l-1}\right\} $$
\begin{equation}
\approx e_{1/q}^{q^4\bar
z_{K-1} z_{K-1}}...e_{1/q}^{q^4\bar z_1 z_1}\exp\left\{
\sum_{l=1}^{K-1} \left( q^4\sum^\infty_{r=0}\frac{q^{2r}(1-q^2)}{1+
q^{2r+4}(1-q^2)\bar z_{K-l}z_{K-l}}\right)
\bar z_{K-l}\Delta z_{K-l}\right\}  \ .      \label{41}
\end{equation}
Substituting (\ref{41}) into (\ref{40}) and taking the continuous limit
$\Delta t\rightarrow 0$ as usual, we finally obtain the path integral
representation for the evolution operator kernel:
\begin{eqnarray}
U(\bar z,z;t^{\prime\prime}-t^\prime)= \int\left(\prod_t d\bar z(t)dz(t)
e_q^{(q^4-1)\bar z(t)z(t)}\right) e_{1/q}^{q^4\bar
z(t^{\prime\prime}) z(t^{\prime\prime})}  \nonumber \\
\times \exp\left\{-\int^{t^{\prime\prime}}_{t^\prime}\left( \phi(\bar
z(t)z(t)) \bar z(t)\dot z(t)+iH_{eff}(\bar z(t)z(t))\right) dt\right\}
\ ,                                                    \label{42}
\end{eqnarray}
where the dot means time derivative and
$$
\phi(\bar z(t)z(t))=\sum^\infty_{r=0}\frac{q^{2r}}{(q^4(1-q^2))^{-1}+
q^{2r}\bar zz}\ .                           $$
Note that in the $q^2\rightarrow 1$ limit
$$
\phi\rightarrow 1\ ,           $$
$$
H_{eff}\rightarrow H_{cl}\equiv \omega\bar z z\ ,    $$
so that in this limit one obtains usual expression for the harmonic
oscillator path integral (cf., e.g., \cite{FS}).
\vskip 1cm

\section{Conclusion}

\hspace{.3in}1) Inhomogeneous Euclidean q-group fixes the differential
calculus on the
complex q-plane and permits to develop Bargmann-Fock representation
for raising and lowering q-operators. The two-dimensional example
shows that inhomogeneous groups strongly restrict possible
differential calculi; the higher dimensional cases deserve further
consideration.

2) We chose q-commuting variables $\bar z(t),z(t)$ on the different
time slices so that their commutation relations do not depend on the
slices they belong to. Such choice was made because of physical
reasons which were not directly related to the possibility of path
integral representation. But this choice is proved to be very crucial
for construction of the q-path integral since it leads to commuting
arguments of exponents in convolution formula (\ref{40}) and permits
to use summation theorems for usual and q-exponents. For example, if
we would use commuting variables on different q-planes as in \cite{BF}
we  would  have  non-commuting  arguments  of  exponentials  for
neighbouring
slices:
$$
(\bar z_iz_{i-1})(\bar z_{i-1}z_{i-2})=
q^2(\bar z_{i-1}z_{i-2})(\bar z_iz_{i-1})    \ .        $$
 From the other hand, the arguments for remote slices
would commute. This makes the construction of action with
definite commutation relation and q-deformed path integral impossible.

3) A natural question arises: is it possibile to take
quasi-classical limit and to derive classical action (and hence
equation of motion) for q-oscillator from our path integral?
Unfortunately, the answer is negative. Indeed, let us restore the  Planck
constant $\hbar$ in expression (\ref{42}) for the path integral,
$$
U(\bar z,z;t^{\prime\prime}-t^\prime)= \int\left(\prod_t\bar z(t)dz(t)
e_q^{(q^4-1)\bar z(t)z(t)/\hbar}\right) e_{1/q}^{q^4\bar
z(t^{\prime\prime}) z(t^{\prime\prime})/\hbar} $$
$$
\times \exp\left\{-\frac{1}{\hbar}\int^{t^{\prime\prime}}_{t^\prime}\left(
\phi(
   \bar
z(t)z(t))\right) \bar z(t)\dot z(t)+iH_{eff}(\bar z(t)z(t))dt\right\}\
,                                                             $$
$$
\phi(\bar z(t)z(t))=\sum^\infty_{r=0}\frac{q^{2r}}{(q^4(1-q^2))^{-1}+
q^{2r}\bar zz/\hbar}\ ,                           $$
$$
H_{eff}=  q^4e_q^{-q^4(1-q^2)\bar zz/\hbar}\omega\bar z z  \ .  $$
One can see that the $\hbar\rightarrow 0$ limit is meaningless for
$q^2\neq 1$. From the other hand, $q^2\rightarrow 1$ limit leads to
usual path integral for harmonic oscillator. Then one can use the
quasi-classical approximation and derive classical equation of motion
in the usual way.

4) We have presented the possibility of  q-deformed path integral but
have not given a recipe for its calculation. This important
problem requires further investigation.
\vskip 10 mm
\centerline{{\bf Acknowledgements}}
We are grateful to P.Pre{\v s}najder for valuable discussions.
A.P.D. thanks  the Research Institute for Theoretical
Physics, University of Helsinki, where part of this work was done,
for hospitality.

\end{document}